\documentclass[a4paper,english,aip, apl, twocolumn, reprint]{revtex4-1}
\pdfoutput=1
\usepackage{amsmath}
\usepackage{bm}
\usepackage{graphicx}
\usepackage{listing}
\usepackage{mathtools} 

\newcommand{\LPTXT}{${\mathrm{L1}}_0$ }
\newcommand{\MUS}{\mu_{\mathrm s}}

\newcommand{\TC}{{T_{\mathrm C}}}
\newcommand{\ABINITIO}{\emph{ab-initio} }
\newcommand{\ETAL}{\emph{et al.}}
\usepackage{color}

\begin{document}

\title{Switching times of nanoscale FePt: finite size effects on linear reversal mechanism}
\author{M. O. A. Ellis, R. W. Chantrell}
\affiliation{Department of Physics, University of York, York, YO10 5DD, UK}
\date{\today}

\begin{abstract}
The linear reversal mechanism in FePt grains ranging from 2.316 nm to 5.404 nm has been
simulated using atomistic spin dynamics, parametrized from \ABINITIO calculations.
The Curie temperature and the critical temperature ($T^*$), at which the linear reversal mechanism occurs, are observed
to decrease with system size whilst the temperature window $ T^* < T < \TC$ increases.
The reversal paths close to the Curie temperature have been calculated, showing 
that for decreasing system size the reversal path becomes more elliptic at lower temperatures, consistent with the decrease
in the Curie temperature arising from finite size effects.
Calculations of the minimum pulse duration show faster switching in small grains and is qualitatively
described by the Landau-Lifshitz-Bloch equation with finite size atomistic parameterization, which
suggests that multiscale modeling of FePt down to a grain size of $\approx 3.5$ nm is possible.
\end{abstract}

\maketitle

The properties of \LPTXT FePt are of paramount importance for the next generation of high density magnetic recording technology, in particular Heat Assisted Magnetic Recording (HAMR). \LPTXT FePt exhibits a very large uniaxial anisotropy of $ K_u \approx 7\times10^6 \text{ Jm}^{-3}$ which allows possible stable grain sizes down to a limit of around $3$ nm.\cite{Weller2000}  To record data the medium is heated close to or above the Curie temperature where the anisotropy is significantly reduced, therefore an understanding of the dynamical behavior of the nanometer FePt grains at elevated temperatures is essential. In particular, at temperatures close to $\TC$ it has been shown that the magnetization reverses by a linear longitudinal route rather than a precessional one.\cite{Kazantseva2009, Barker2010} 
Using the Landau-Lifshitz-Bloch model Kazantseva \emph{et al.}\cite{Kazantseva2009} predicted that the linear reversal regime
occurs at a critical temperature $T^*$ and derived a expression for the reversal time in this regime. Following this Barker \emph{et al.}\cite{Barker2010} observed the linear reversal mechanism using an atomistic model of FePt which agreed well with the LLB modeling.  However the question of linear reversal in finite nanometer grains and the size dependence of the switching properties remains open.

Here an atomistic model is used to investigate the properties of nanometer grains from 2.316 nm to 5.404 nm at temperatures close to the Curie temperature. Using Langevin dynamics methods we investigate the static and dynamic properties as a function of the grain size. 
We calculate the equilibrium magnetization and susceptibilities for each system size which allow us to parametrize the LLB 
model. Using this parametrization we can then compare the reversal path ellipticity and times calculated using the atomistic model to those predicted by the LLB.

The properties of nanometer FePt particles have been investigated using \ABINITIO methods by Chepulskii and Butler.\cite{Chepulskii2012} Their calculations show that
the magneto-crystalline anisotropy and magnetic moment varies significantly with system size and shape. However, the sizes considered are smaller than 1 nm so here we use the parameterization of bulk materials given by Mryasov \emph{et al.}\cite{Mryasov2005} 
The finite size effects in FePt have been investigated using a similar model by Lyberatos \ETAL\cite{Lyberatos2013} and Hovorka \ETAL\cite{Hovorka2012} Lyberatos \emph{et al.} have further investigated the grain size effects on the HAMR write process showing that smaller grain sizes require larger write fields.\cite{Lyberatos2014}

Atomistic spin models are increasingly used for dynamic magnetization calculations (Evans \ETAL\cite{Evans2014}). 
The \emph{ab-intio} simulations of Mryasov \emph{et al.} show that the \LPTXT FePt spin
Hamiltonian can be written in the following form, involving only Fe degrees of freedom with parameters mediated by the Pt:
\begin{equation}
\mathcal{H} = -\sum_{i\neq j} J_{ij} \mathbf{S}_i \cdot \mathbf{S}_j - \sum_i d_i^{(0)} (S_i^z)^2 - \sum_{i\neq j} d_{ij}^{(2)} S_i^z S_j^z
\end{equation}
The first term is the mediated exchange interaction, the second term a uniaxial anisotropy and the third a two-ion anisotropy which represents an anisotropic exchange interaction. The sum $j$ is over the atoms within the interaction range, which extends for approximately 5 unit cells. It is important to note that our model includes a significant finite size effect in that, as noted by Nowak \emph{et al.}\cite{Nowak2005}, there is a loss of anisotropy at the surface due to the reduction of coordination which results in a reduction of the 2-ion anisotropy contribution. These finite size effects mean the total anisotropy varies from $K_u= 8.3 \times 10^{6} \text{ Jm}^{-3}$ for a 4.632 nm particle to $K_u = 7.3 \times 10^{6} \text{ Jm}^{-3}$ for a 2.316 nm particle. The dipolar field was not included in the 
calculations as at temperatures close to $\TC$ it has negligible effect and is not important in the
linear reversal mechanism.

The time evolution of the magnetic properties is calculated using the Langevin dynamic approach. The LLG equation for an atomic spin is:
\begin{equation}
\frac{ \partial \mathbf{S}_i}{\partial t} = \frac{- \gamma}{ (1 + \lambda^2) \mu_s} \mathbf{S}_i \times  \left( \mathbf{H}_i + \lambda \mathbf{S}_i \times \mathbf{H}_i \right)
\end{equation}
Where $\gamma = 1.76\times 10^{11} \text{s}^{-1}\text{T}^{-1}$ is the gyromagnetic ratio, $\MUS = 3.23 \mu_B$ is the saturation magnetic moment of the combined Fe and Pt atoms, $\lambda = 0.1$ is the atomistic damping/coupling parameter and $\mathbf{H}_i$ is
the effective field acting on the spin. For these parameters the saturation magnetization is $M_s = 1.075\times10^6 \text{ JT}^{-1}\text{m}^{-3}$.
Temperature is introduced via a thermal noise term which is added to the effective field, $\mathbf{H}_i = -\partial \mathcal{H}/\partial \mathbf{S}_i + \boldsymbol{ \xi}_i$.
The thermal noise is a Gaussian stochastic process with the following mean and variance.\cite{Brown1963}
\begin{align}
\langle \xi_{i,\alpha}(t) \rangle & = 0 \\
\langle \xi_{i,\alpha}(t) \xi_{j,\beta}(t') \rangle & = \delta_{ij}\delta_{\alpha\beta}\delta(t-t') \frac{ 2 \mu_s \lambda k_B T}{\gamma}
\end{align}
where i,j refer to atom numbers, $\alpha,\beta $ Cartesian coordinates and $t,t'$ the time. 
The resulting stochastic LLG equation is numerically solved using the Semi-Implicit
integration scheme\cite{Mentink2010, Ellis2012} with a time step $\Delta t = 10^{-16}$ s which has
been tested to show stability.

\begin{figure}[tbp]
    \centering
    \includegraphics[width=\columnwidth]{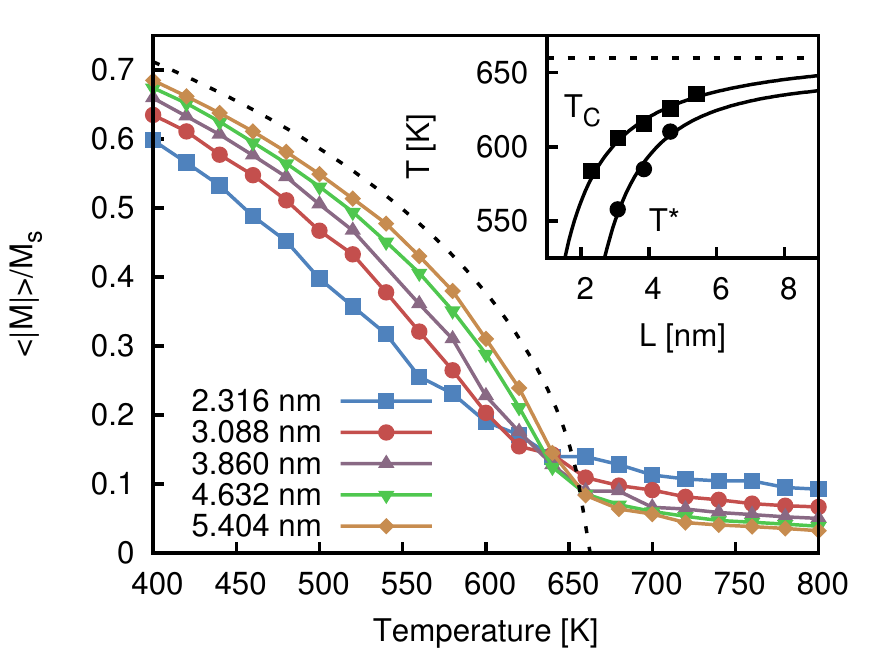}
    \caption{
    The temperature dependence of the magnetization for particle sizes of 2.316 nm to 5.404 nm. The dashed black line shows the magnetization for a bulk system from ref.~\onlinecite{Kazantseva2008PRB} for comparison. As the system size decreases the magnetization below the Curie temperature decreases while above it the magnetization
    increases. The inset shows the Curie temperatures which have been extracted from fitting and $T^*$ extracted from figure \ref{fig:LD_suscept_ratio}. The solid lines are equation \ref{eqn:scaling_TC} fitted to find the critical exponent.
    }
    \label{fig:MvT}
\end{figure}

We first investigate the temperature dependence of the magnetization and the apparent Curie temperature. 
The results are shown in figure \ref{fig:MvT} for a selection of system sizes. Clearly the smaller systems have a reduced equilibrium magnetization at a given temperature, as expected as a result of reduced co-ordination at the surfaces. The apparent Curie temperature can be extracted from fitting $m(T) = (1 - T/T_C(L))^\beta$ to the magnetization curves.\cite{Lyberatos2013} The extracted $T_C$ values are reduced significantly from the bulk and described well by finite system size scaling theory:\cite{Hovorka2012} 
\begin{align}
 \frac{ T_C^\infty - T_C}{T_C^\infty} = \left( \frac{L}{d_0} \right)^{-1/\nu} 
\label{eqn:scaling_TC}
\end{align}
The inset in figure \ref{fig:MvT} shows the fitting to obtain $T^\infty_C = 665$ K, $\nu = 0.85699, d_0 = 0.4039$ nm which agree well with refs.~\onlinecite{Hovorka2012} and \onlinecite{Lyberatos2013} for the long range FePt Hamiltonian which we utilize.
\begin{figure}[tbp!]
    \centering
    \includegraphics[width=\columnwidth]{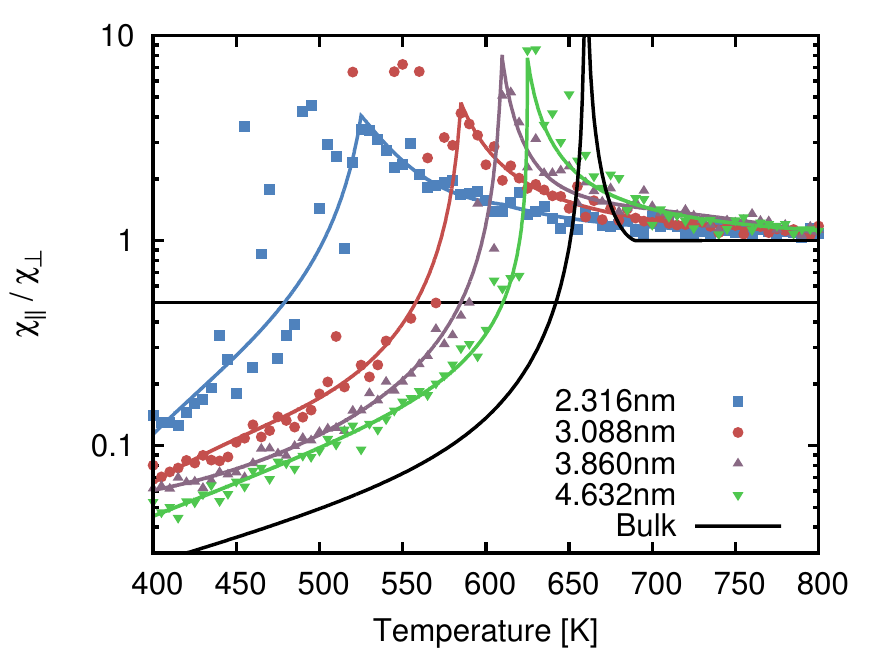}
    \caption{ The ratio of the parallel and perpendicular susceptibilities reach a peak at different temperatures consistent with the Curie temperature extracted from the magnetization. The solid lines show functions that have been fitted the susceptibilities and the horizontal solid line gives $\tilde{\chi}_\parallel/\tilde{\chi}_\perp = 1/2 $ defining the transition to linear reversal.
    }
    \label{fig:LD_suscept_ratio}
\end{figure}

The switching time at elevated temperatures is strongly affected by the reversal mechanism, which is determined by the ratio of longitudinal and transverse susceptibilities, $\tilde{\chi}_\parallel, \tilde{\chi}_\perp$. Consequently we next investigate the ratio of susceptibilities and make a comparison with bulk properties using an expression for the free energy, in the absence of an externally applied field~\cite{Garanin1997,Evans2012}:
\begin{equation}
    \frac{F}{ M_s V}  = 
	\begin{dcases}
		\frac{m_x^2 + m_y^2}{2 \tilde{\chi}_\perp} + \frac{ \left( m^2 - m_e^2 \right)^2 }{8 m_e^2 \tilde{\chi}_\parallel} &  T < T_c \\
                     \frac{m_x^2 + m_y^2}{2 \tilde{\chi}_\perp} 
					 + \frac{3}{20\tilde{\chi}_\parallel} \frac{T_C}{T-T_C} \\
					\quad \quad \times \left( m^2 + \frac{5}{3} \frac{T-T_C}{T_C} \right)^2  &  T > T_c  ,
	\end{dcases}		
\label{free_energy}
\end{equation}
where $m$ is the reduced magnetization, $m_e$ the equilibrium magnetization and $\tilde{\chi}_\parallel, \tilde{\chi}_\perp$ are respectively the reduced parallel and
perpendicular susceptibility. We note that eqn. \ref{free_energy} gives the free energy governing the behavior of the LLB equation derived in ref.~\onlinecite{Garanin1997}, which allows us to connect the current calculations to the LLB equation as will be shown later.
From the free energy we find the following expression for the orientation dependence of the magnetization length, $m(\theta)$ for a bulk material:
\begin{align}
    m(\theta) = m_e \left( 1 - 2 \frac{ \tilde{ \chi}_\parallel }{ \tilde{ \chi}_\perp} \sin^2(\theta) \right)^{\frac{1}{2}}
    \label{m_vs_theta}
\end{align}
Eqn. \ref{m_vs_theta} gives an important relationship between the ellipticity parameter and the characteristic parameters of the LLB equation. Specifically,  the ratio of the longitudinal and perpendicular susceptibility determines the ellipticity of the reversal path. Eqn. \ref{m_vs_theta} predicts a critical temperature, $T^*$, corresponding to
$ \tilde{ \chi}_\parallel(T^*) / \tilde{ \chi}_\perp(T^*) = 1/2$, at which point the perpendicular magnetization vanishes. $T^*$ is the critical temperature above which the magnetization will reverse through the linear switching mechanism.
By calculating the equilibrium magnetization and susceptibilities for finite nanometer systems equations \ref{free_energy} and \ref{m_vs_theta} can be parametrized.
The longitudinal and transverse susceptibilities have been calculated as $\chi_z$ and $\chi_x + \chi_y$ respectively. The ratio $ \tilde{ \chi}_\parallel(T) / \tilde{ \chi}_\perp(T)$ is shown in fig ~\ref{fig:LD_suscept_ratio}. The points show the susceptibilities from the atomistic simulations while the solid lines show a polynomial function that has been fitted to inverse susceptibilities individually in a similar manner to ref.~\onlinecite{Kazantseva2008PRB}. The significant noise in the small grains arises from the thermal switching of the magnetization during the averaging of $m_z$ leading to ergodicity breaking.\cite{Binder1997} As the grain size decreases the critical temperature $T^*$  decreases and the apparent temperature window $ T^* < T < \TC$ of the linear reversal region widens. $T^*$ is shown in the
inset of fig \ref{fig:MvT} alongside $T_C$ which highlights the widening temperature window.
Although the free energy expression given in eqn. \ref{free_energy} is strictly valid only for bulk systems, the temperature variation of $ \tilde{ \chi}_\parallel(T) / \tilde{ \chi}_\perp(T)$ for all system sizes is consistent with the behavior shown by the reversal paths and the ellipticity.

\begin{figure*}[tb]
    \centering
    \includegraphics[width=\textwidth]{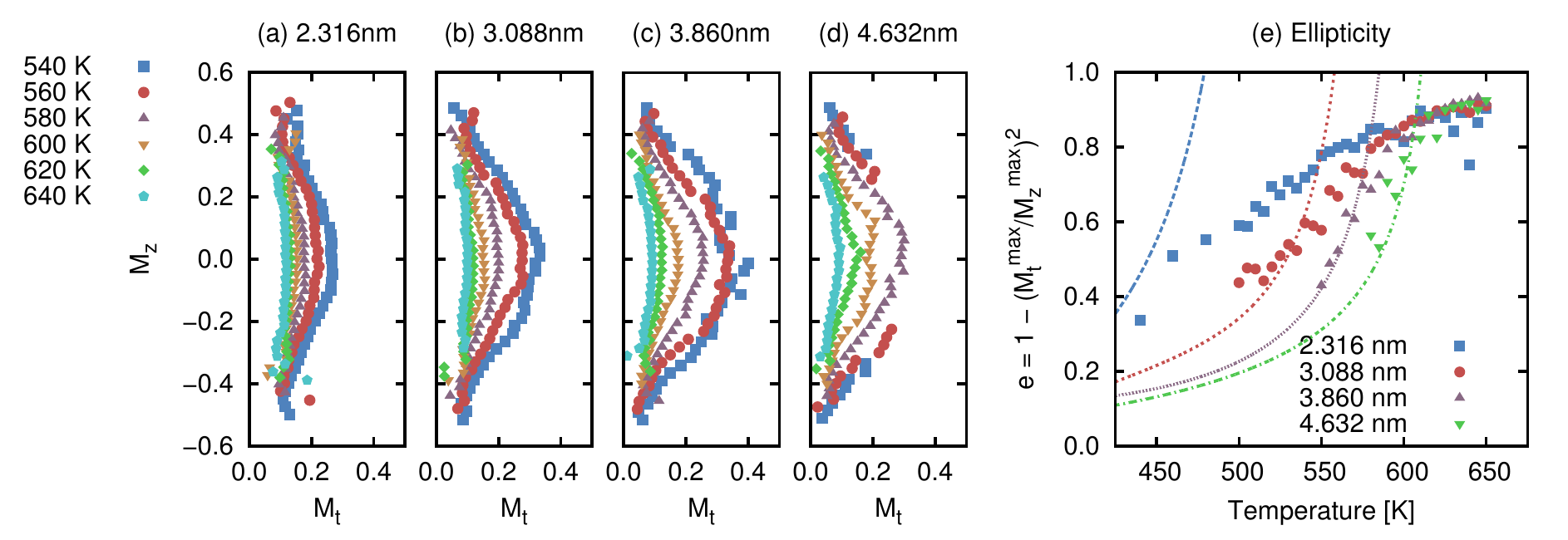}
    \caption{ The reversal paths of (a) 2.316 nm, (b) 3.088 nm, (c) 3.860 nm and (d) 4.632 nm systems using a reversing field of 1T. The reversal
    paths are calculated at different temperatures slightly below the Curie temperature. For the larger system sizes the
    lower temperatures have not reversed with in the simulation time limit but as the system size decreases at the same temperature
    the system now reverses within the given time.
    (e) The ellipticity extracted from the reversal paths in (a)-(d) using equation \ref{eqn:ellip}. The dashed lines show the ellipticity that is expected by the LLB model using the finite size susceptibilities shown in figure \ref{fig:LD_suscept_ratio}.
    }
    \label{fig:reversal_path_1T}
\end{figure*}

Figure \ref{fig:reversal_path_1T}.(a)-(d) shows the reversal paths for system size 2.314 nm to 4.632 nm using a 1 T field to investigate the temperature dependence of the reversal path. In the 4.632 nm particle the magnetization does not reverse
within the simulation time limit of 400 ps for 540 K and 560 K but then there is a sharp transition to linear reversal, characterized by vanishing transverse magnetization, at 640K. The smaller systems reverse at 
lower temperatures but there is a less critical transition to the linear reversal path.

To further investigate the effect of the system size on the transition to the linear regime, we calculate the ellipticity of the reversal paths, defined as
\begin{equation}
    e = 1 - \left( \frac{M_t^{\text{max}} }{M_z^{\text{max}} }  \right)^2
	\label{eqn:ellip}
\end{equation}
The ellipticity is shown in figure \ref{fig:reversal_path_1T}.(e) for the reversal paths shown in fig. \ref{fig:reversal_path_1T}.(a)-(d). As discussed previously the transition to the linear
regime (where $ e \approx 1$) occurs within a smaller temperature window for the larger systems. Also since the Curie temperature decreases
for smaller system sizes the window is centered at lower temperatures. The lines shown in fig \ref{fig:reversal_path_1T}.(e) show the ellipticity from the LLB model which is $e_\text{LLB} = 2 \tilde{ \chi}_\parallel(T) / \tilde{ \chi}_\perp(T)$, this shows that 
the LLB model reasonably predicts the ellipticity but significant deviations occur for the smaller particles.

 In terms of HAMR, the temperature dependence of the relaxation time is an important factor. It has been shown by Evans {\emph et al.}~\cite{Evans2012a} that the achievable recording density in HAMR is set by the equilibrium magnetization at the temperature at which the magnetization freezes. Consequently, it is vital to the HAMR process for the magnetization to be as close as possible to the equilibrium value at a given temperature, emphasizing the importance of the linear reversal mechanism and its associated fast switching time governed by longitudinal rather than transverse relaxation.
\begin{figure}[tb!]
    \centering
    \includegraphics[width=\columnwidth]{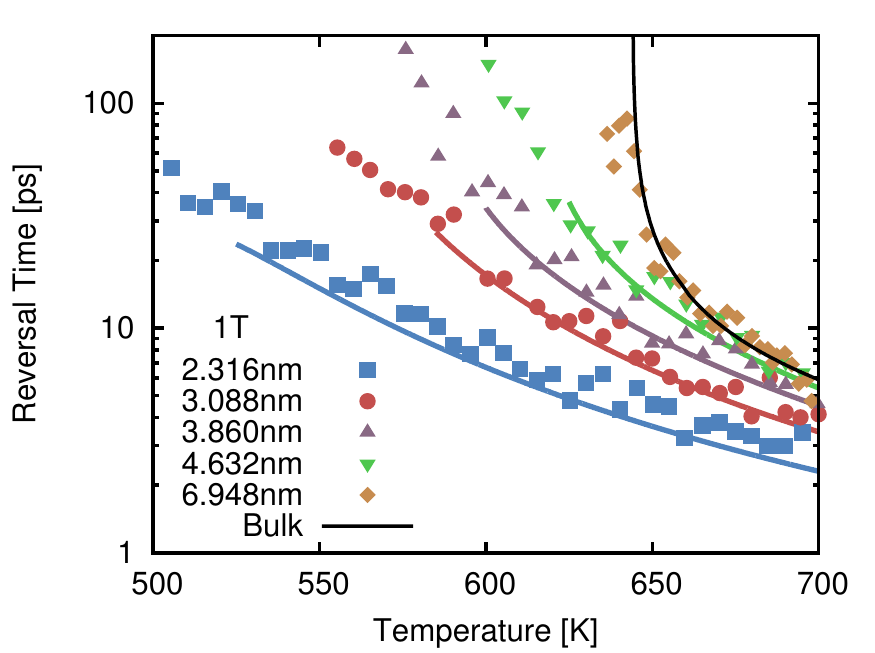}
    \caption{
   	The reversal time in a constant 1 Tesla reversing field. The solid lines show the analytic minimum pulse duration which is derived in ref.~\onlinecite{Kazantseva2009}
    the parameters for which are calculated from our atomistic simulations for the finite systems while the black line is for bulk.
    }
    \label{fig:rev_time}
\end{figure}
As a measure of the relaxation time we calculate the minimal pulse duration\cite{Barker2010}; this is the time taken for $m_z$ to first pass the $m_z = 0$ plane starting from an initial fully magnetized state. Figure \ref{fig:rev_time} shows the minimal pulse duration calculated with a 1 T reversing field relevant to the HAMR process. 
The results show that there is a significant reduction in the relaxation time in small grains which arises from a convolution of two factors, the decrease of intrinsic properties and the transition from circular to linear reversal. The intrinsic parameters $K$ and $M_s$ both decrease with temperature and depend strongly on the system size. Since a large part of the anisotropy arises from the 2-ion interaction there is a strong decrease with grain size and thus the relaxation time predictable from the Arrhenius-N\'{e}el law.\cite{Neel1949} In the elliptical and linear regimes the free energy barrier is reduced significantly relative to the coherent reversal mechanism, leading to a further reduction of switching time with temperature over and above that predicted by the Arrhenius-N\'{e}el law. We note that both factors are essentially thermodynamic and emerge naturally from the atomistic model.

We compare these results to the expression for the minimum pulse duration derived from the LLB equation by Kazantseva \emph{et al.}\cite{Kazantseva2009} The minimum pulse duration depends on the equilibrium magnetization and susceptibilities so using the values already presented it can be computed for each specific grain size, which are shown as solid lines in figure \ref{fig:rev_time}. There is a good agreement but since the free energy assumes an infinite system the problems arise at size dependent $\TC$ so only the section above is shown.
The analytic expression with finite input parameters shows qualitative agreement with the atomistic reversal times showing that for moderate system sizes the LLB equation is applicable using parameters determined for the specific system size. The reversal times for the 6.948 nm particle are similar to the bulk LLB curve. For the smallest system sizes the validity is questionable since it is far from the bulk regime in which the LLB is derived. However, for the sizes in excess of 3.5 nm potentially usable as HAMR media, the parameterization seems a practical proposition.

In summary we have investigated the effects of finite size on the linear reversal regime of FePt which is of
technological importance for HAMR. Atomistic calculations of the temperature
dependence of the magnetization shows that $\TC$ decreases with decreasing system size
as does the critical temperature for the linear reversal regime, $T^*$ determined from the susceptibilities. The transition to linear reversal also takes place at a lower temperature, with the temperature range of linear reversal increasing with decreasing size.
 The reversal characteristics
have been calculated showing that the reversal paths do exhibit a trend towards
linear reversal at $T^*$ but in smaller grains the criticality of the transition is reduced.
We find that the expression for the reversal time derived from the LLB equation gives switching times close to the atomistic results using values appropriate for each individual nanoparticle diameter. 
This suggests that a multiscale approach for HAMR media modeling is feasible using the LLB equation as a macrospin approximation to the behavior of nanoparticles, with parameters determined for each grain size. Given the importance of linear reversal to the HAMR process, this represents an important result for simulations of future generations of HAMR media as the grain size is inevitably reduced as linear densities increase.

The authors would like to thank A. Lyberatos for helpful discussions. Financial support  from the Advanced Storage Technology Consortium is gratefully acknowledged.

\bibliography{references}

\begin{thebibliography}{19}%
\makeatletter
\providecommand \@ifxundefined [1]{%
 \@ifx{#1\undefined}
}%
\providecommand \@ifnum [1]{%
 \ifnum #1\expandafter \@firstoftwo
 \else \expandafter \@secondoftwo
 \fi
}%
\providecommand \@ifx [1]{%
 \ifx #1\expandafter \@firstoftwo
 \else \expandafter \@secondoftwo
 \fi
}%
\providecommand \natexlab [1]{#1}%
\providecommand \enquote  [1]{``#1''}%
\providecommand \bibnamefont  [1]{#1}%
\providecommand \bibfnamefont [1]{#1}%
\providecommand \citenamefont [1]{#1}%
\providecommand \href@noop [0]{\@secondoftwo}%
\providecommand \href [0]{\begingroup \@sanitize@url \@href}%
\providecommand \@href[1]{\@@startlink{#1}\@@href}%
\providecommand \@@href[1]{\endgroup#1\@@endlink}%
\providecommand \@sanitize@url [0]{\catcode `\\12\catcode `\$12\catcode
  `\&12\catcode `\#12\catcode `\^12\catcode `\_12\catcode `\%12\relax}%
\providecommand \@@startlink[1]{}%
\providecommand \@@endlink[0]{}%
\providecommand \url  [0]{\begingroup\@sanitize@url \@url }%
\providecommand \@url [1]{\endgroup\@href {#1}{\urlprefix }}%
\providecommand \urlprefix  [0]{URL }%
\providecommand \Eprint [0]{\href }%
\providecommand \doibase [0]{http://dx.doi.org/}%
\providecommand \selectlanguage [0]{\@gobble}%
\providecommand \bibinfo  [0]{\@secondoftwo}%
\providecommand \bibfield  [0]{\@secondoftwo}%
\providecommand \translation [1]{[#1]}%
\providecommand \BibitemOpen [0]{}%
\providecommand \bibitemStop [0]{}%
\providecommand \bibitemNoStop [0]{.\EOS\space}%
\providecommand \EOS [0]{\spacefactor3000\relax}%
\providecommand \BibitemShut  [1]{\csname bibitem#1\endcsname}%
\let\auto@bib@innerbib\@empty
\bibitem [{\citenamefont {Weller}\ \emph {et~al.}(2000)\citenamefont {Weller},
  \citenamefont {Moser}, \citenamefont {Folks},\ and\ \citenamefont
  {Best}}]{Weller2000}%
  \BibitemOpen
  \bibfield  {author} {\bibinfo {author} {\bibfnamefont {D.}~\bibnamefont
  {Weller}}, \bibinfo {author} {\bibfnamefont {A.}~\bibnamefont {Moser}},
  \bibinfo {author} {\bibfnamefont {L.}~\bibnamefont {Folks}}, \ and\ \bibinfo
  {author} {\bibfnamefont {M.}~\bibnamefont {Best}},\ }\href
  {http://ieeexplore.ieee.org/xpls/abs\_all.jsp?arnumber=824418} {\bibfield
  {journal} {\bibinfo  {journal} {IEEE Trans. Magn.}\ }\textbf {\bibinfo
  {volume} {36}},\ \bibinfo {pages} {1} (\bibinfo {year} {2000})}\BibitemShut
  {NoStop}%
\bibitem [{\citenamefont {Kazantseva}\ \emph {et~al.}(2009)\citenamefont
  {Kazantseva}, \citenamefont {Hinzke}, \citenamefont {Chantrell},\ and\
  \citenamefont {Nowak}}]{Kazantseva2009}%
  \BibitemOpen
  \bibfield  {author} {\bibinfo {author} {\bibfnamefont {N.}~\bibnamefont
  {Kazantseva}}, \bibinfo {author} {\bibfnamefont {D.}~\bibnamefont {Hinzke}},
  \bibinfo {author} {\bibfnamefont {R.~W.}\ \bibnamefont {Chantrell}}, \ and\
  \bibinfo {author} {\bibfnamefont {U.}~\bibnamefont {Nowak}},\ }\href
  {\doibase 10.1209/0295-5075/86/27006} {\bibfield  {journal} {\bibinfo
  {journal} {Europhys. Lett.}\ }\textbf {\bibinfo {volume} {86}},\ \bibinfo
  {pages} {27006} (\bibinfo {year} {2009})}\BibitemShut {NoStop}%
\bibitem [{\citenamefont {Barker}\ \emph {et~al.}(2010)\citenamefont {Barker},
  \citenamefont {Evans}, \citenamefont {Chantrell}, \citenamefont {Hinzke},\
  and\ \citenamefont {Nowak}}]{Barker2010}%
  \BibitemOpen
  \bibfield  {author} {\bibinfo {author} {\bibfnamefont {J.}~\bibnamefont
  {Barker}}, \bibinfo {author} {\bibfnamefont {R.~F.~L.}\ \bibnamefont
  {Evans}}, \bibinfo {author} {\bibfnamefont {R.~W.}\ \bibnamefont
  {Chantrell}}, \bibinfo {author} {\bibfnamefont {D.}~\bibnamefont {Hinzke}}, \
  and\ \bibinfo {author} {\bibfnamefont {U.}~\bibnamefont {Nowak}},\ }\href
  {\doibase 10.1063/1.3515928} {\bibfield  {journal} {\bibinfo  {journal}
  {Appl. Phys. Lett.}\ }\textbf {\bibinfo {volume} {97}},\ \bibinfo {pages}
  {192504} (\bibinfo {year} {2010})}\BibitemShut {NoStop}%
\bibitem [{\citenamefont {Chepulskii}\ and\ \citenamefont
  {Butler}(2012)}]{Chepulskii2012}%
  \BibitemOpen
  \bibfield  {author} {\bibinfo {author} {\bibfnamefont {R.~V.}\ \bibnamefont
  {Chepulskii}}\ and\ \bibinfo {author} {\bibfnamefont {W.~H.}\ \bibnamefont
  {Butler}},\ }\href {\doibase 10.1063/1.3700746} {\bibfield  {journal}
  {\bibinfo  {journal} {Appl. Phys. Lett.}\ }\textbf {\bibinfo {volume}
  {100}},\ \bibinfo {pages} {142405} (\bibinfo {year} {2012})}\BibitemShut
  {NoStop}%
\bibitem [{\citenamefont {Mryasov}\ \emph {et~al.}(2005)\citenamefont
  {Mryasov}, \citenamefont {Nowak}, \citenamefont {Chantrell},\ and\
  \citenamefont {Guslienko}}]{Mryasov2005}%
  \BibitemOpen
  \bibfield  {author} {\bibinfo {author} {\bibfnamefont {O.~N.}\ \bibnamefont
  {Mryasov}}, \bibinfo {author} {\bibfnamefont {U.}~\bibnamefont {Nowak}},
  \bibinfo {author} {\bibfnamefont {R.~W.}\ \bibnamefont {Chantrell}}, \ and\
  \bibinfo {author} {\bibfnamefont {K.~Y.}\ \bibnamefont {Guslienko}},\ }\href
  {\doibase 10.1209/epl/i2004-10404-2} {\bibfield  {journal} {\bibinfo
  {journal} {Europhys. Lett.}\ }\textbf {\bibinfo {volume} {69}},\ \bibinfo
  {pages} {805} (\bibinfo {year} {2005})}\BibitemShut {NoStop}%
\bibitem [{\citenamefont {Lyberatos}\ \emph {et~al.}(2013)\citenamefont
  {Lyberatos}, \citenamefont {Weller},\ and\ \citenamefont
  {Parker}}]{Lyberatos2013}%
  \BibitemOpen
  \bibfield  {author} {\bibinfo {author} {\bibfnamefont {A.}~\bibnamefont
  {Lyberatos}}, \bibinfo {author} {\bibfnamefont {D.}~\bibnamefont {Weller}}, \
  and\ \bibinfo {author} {\bibfnamefont {G.~J.}\ \bibnamefont {Parker}},\
  }\href {\doibase 10.1063/1.4839875} {\bibfield  {journal} {\bibinfo
  {journal} {J. Appl. Phys.}\ }\textbf {\bibinfo {volume} {114}},\ \bibinfo
  {pages} {233904} (\bibinfo {year} {2013})}\BibitemShut {NoStop}%
\bibitem [{\citenamefont {Hovorka}\ \emph {et~al.}(2012)\citenamefont
  {Hovorka}, \citenamefont {Devos}, \citenamefont {Coopman}, \citenamefont
  {Fan}, \citenamefont {Aas}, \citenamefont {Evans},\ and\ \citenamefont
  {Chen}}]{Hovorka2012}%
  \BibitemOpen
  \bibfield  {author} {\bibinfo {author} {\bibfnamefont {O.}~\bibnamefont
  {Hovorka}}, \bibinfo {author} {\bibfnamefont {S.}~\bibnamefont {Devos}},
  \bibinfo {author} {\bibfnamefont {Q.}~\bibnamefont {Coopman}}, \bibinfo
  {author} {\bibfnamefont {W.~J.}\ \bibnamefont {Fan}}, \bibinfo {author}
  {\bibfnamefont {C.~J.}\ \bibnamefont {Aas}}, \bibinfo {author} {\bibfnamefont
  {R.~F.~L.}\ \bibnamefont {Evans}}, \ and\ \bibinfo {author} {\bibfnamefont
  {X.}~\bibnamefont {Chen}},\ }\href {\doibase 10.1063/1.4740075} {\bibfield
  {journal} {\bibinfo  {journal} {Appl. Phys. Lett.}\ }\textbf {\bibinfo
  {volume} {101}},\ \bibinfo {pages} {052406} (\bibinfo {year}
  {2012})}\BibitemShut {NoStop}%
\bibitem [{\citenamefont {Lyberatos}\ \emph {et~al.}(2014)\citenamefont
  {Lyberatos}, \citenamefont {Weller},\ and\ \citenamefont
  {Parker}}]{Lyberatos2014}%
  \BibitemOpen
  \bibfield  {author} {\bibinfo {author} {\bibfnamefont {A.}~\bibnamefont
  {Lyberatos}}, \bibinfo {author} {\bibfnamefont {D.}~\bibnamefont {Weller}}, \
  and\ \bibinfo {author} {\bibfnamefont {G.~J.}\ \bibnamefont {Parker}},\
  }\href@noop {} {\bibfield  {journal} {\bibinfo  {journal} {IEEE Trans.
  Magn.}\ }\textbf {\bibinfo {volume} {50}} (\bibinfo {year}
  {2014})}\BibitemShut {NoStop}%
\bibitem [{\citenamefont {Evans}\ \emph {et~al.}(2014)\citenamefont {Evans},
  \citenamefont {Fan}, \citenamefont {Chureemart}, \citenamefont {Ostler},
  \citenamefont {Ellis},\ and\ \citenamefont {Chantrell}}]{Evans2014}%
  \BibitemOpen
  \bibfield  {author} {\bibinfo {author} {\bibfnamefont {R.~F.~L.}\
  \bibnamefont {Evans}}, \bibinfo {author} {\bibfnamefont {W.~J.}\ \bibnamefont
  {Fan}}, \bibinfo {author} {\bibfnamefont {P.}~\bibnamefont {Chureemart}},
  \bibinfo {author} {\bibfnamefont {T.~A.}\ \bibnamefont {Ostler}}, \bibinfo
  {author} {\bibfnamefont {M.~O.~A.}\ \bibnamefont {Ellis}}, \ and\ \bibinfo
  {author} {\bibfnamefont {R.~W.}\ \bibnamefont {Chantrell}},\ }\href {\doibase
  10.1088/0953-8984/26/10/103202} {\bibfield  {journal} {\bibinfo  {journal}
  {J. Phys. Condens. Matter}\ }\textbf {\bibinfo {volume} {26}},\ \bibinfo
  {pages} {103202} (\bibinfo {year} {2014})}\BibitemShut {NoStop}%
\bibitem [{\citenamefont {Nowak}\ \emph {et~al.}(2005)\citenamefont {Nowak},
  \citenamefont {Mryasov}, \citenamefont {Wieser}, \citenamefont {Guslienko},\
  and\ \citenamefont {Chantrell}}]{Nowak2005}%
  \BibitemOpen
  \bibfield  {author} {\bibinfo {author} {\bibfnamefont {U.}~\bibnamefont
  {Nowak}}, \bibinfo {author} {\bibfnamefont {O.}~\bibnamefont {Mryasov}},
  \bibinfo {author} {\bibfnamefont {R.}~\bibnamefont {Wieser}}, \bibinfo
  {author} {\bibfnamefont {K.}~\bibnamefont {Guslienko}}, \ and\ \bibinfo
  {author} {\bibfnamefont {R.}~\bibnamefont {Chantrell}},\ }\href {\doibase
  10.1103/PhysRevB.72.172410} {\bibfield  {journal} {\bibinfo  {journal} {Phys.
  Rev. B}\ }\textbf {\bibinfo {volume} {72}},\ \bibinfo {pages} {172410}
  (\bibinfo {year} {2005})}\BibitemShut {NoStop}%
\bibitem [{\citenamefont {Brown}(1963)}]{Brown1963}%
  \BibitemOpen
  \bibfield  {author} {\bibinfo {author} {\bibfnamefont {W.~F.}\ \bibnamefont
  {Brown}},\ }\href {http://prola.aps.org/abstract/PR/v130/i5/p1677\_1}
  {\bibfield  {journal} {\bibinfo  {journal} {Phys. Rev.}\ }\textbf {\bibinfo
  {volume} {130}},\ \bibinfo {pages} {1677} (\bibinfo {year}
  {1963})}\BibitemShut {NoStop}%
\bibitem [{\citenamefont {Mentink}\ \emph {et~al.}(2010)\citenamefont
  {Mentink}, \citenamefont {Tretyakov}, \citenamefont {Fasolino}, \citenamefont
  {Katsnelson},\ and\ \citenamefont {Rasing}}]{Mentink2010}%
  \BibitemOpen
  \bibfield  {author} {\bibinfo {author} {\bibfnamefont {J.~H.}\ \bibnamefont
  {Mentink}}, \bibinfo {author} {\bibfnamefont {M.~V.}\ \bibnamefont
  {Tretyakov}}, \bibinfo {author} {\bibfnamefont {A.}~\bibnamefont {Fasolino}},
  \bibinfo {author} {\bibfnamefont {M.~I.}\ \bibnamefont {Katsnelson}}, \ and\
  \bibinfo {author} {\bibfnamefont {T.}~\bibnamefont {Rasing}},\ }\href
  {http://stacks.iop.org/0953-8984/22/i=17/a=176001} {\bibfield  {journal}
  {\bibinfo  {journal} {J. Phys. Condens. Matter}\ }\textbf {\bibinfo {volume}
  {22}},\ \bibinfo {pages} {176001} (\bibinfo {year} {2010})}\BibitemShut
  {NoStop}%
\bibitem [{\citenamefont {Ellis}\ \emph {et~al.}(2012)\citenamefont {Ellis},
  \citenamefont {Ostler},\ and\ \citenamefont {Chantrell}}]{Ellis2012}%
  \BibitemOpen
  \bibfield  {author} {\bibinfo {author} {\bibfnamefont {M.~O.~A.}\
  \bibnamefont {Ellis}}, \bibinfo {author} {\bibfnamefont {T.~A.}\ \bibnamefont
  {Ostler}}, \ and\ \bibinfo {author} {\bibfnamefont {R.~W.}\ \bibnamefont
  {Chantrell}},\ }\href {\doibase 10.1103/PhysRevB.86.174418} {\bibfield
  {journal} {\bibinfo  {journal} {Phys. Rev. B}\ }\textbf {\bibinfo {volume}
  {86}},\ \bibinfo {pages} {174418} (\bibinfo {year} {2012})}\BibitemShut
  {NoStop}%
\bibitem [{\citenamefont {Kazantseva}\ \emph {et~al.}(2008)\citenamefont
  {Kazantseva}, \citenamefont {Hinzke}, \citenamefont {Nowak}, \citenamefont
  {Chantrell}, \citenamefont {Atxitia},\ and\ \citenamefont
  {Chubykalo-Fesenko}}]{Kazantseva2008PRB}%
  \BibitemOpen
  \bibfield  {author} {\bibinfo {author} {\bibfnamefont {N.}~\bibnamefont
  {Kazantseva}}, \bibinfo {author} {\bibfnamefont {D.}~\bibnamefont {Hinzke}},
  \bibinfo {author} {\bibfnamefont {U.}~\bibnamefont {Nowak}}, \bibinfo
  {author} {\bibfnamefont {R.~W.}\ \bibnamefont {Chantrell}}, \bibinfo {author}
  {\bibfnamefont {U.}~\bibnamefont {Atxitia}}, \ and\ \bibinfo {author}
  {\bibfnamefont {O.}~\bibnamefont {Chubykalo-Fesenko}},\ }\href {\doibase
  10.1103/PhysRevB.77.184428} {\bibfield  {journal} {\bibinfo  {journal} {Phys.
  Rev. B}\ }\textbf {\bibinfo {volume} {77}},\ \bibinfo {pages} {184428}
  (\bibinfo {year} {2008})}\BibitemShut {NoStop}%
\bibitem [{\citenamefont {Garanin}(1997)}]{Garanin1997}%
  \BibitemOpen
  \bibfield  {author} {\bibinfo {author} {\bibfnamefont {D.}~\bibnamefont
  {Garanin}},\ }\href {\doibase 10.1103/PhysRevB.55.3050} {\bibfield  {journal}
  {\bibinfo  {journal} {Phys. Rev. B}\ }\textbf {\bibinfo {volume} {55}},\
  \bibinfo {pages} {3050} (\bibinfo {year} {1997})}\BibitemShut {NoStop}%
\bibitem [{\citenamefont {Evans}\ \emph
  {et~al.}(2012{\natexlab{a}})\citenamefont {Evans}, \citenamefont {Hinzke},
  \citenamefont {Atxitia}, \citenamefont {Nowak}, \citenamefont {Chantrell},\
  and\ \citenamefont {Chubykalo-Fesenko}}]{Evans2012}%
  \BibitemOpen
  \bibfield  {author} {\bibinfo {author} {\bibfnamefont {R.~F.}\ \bibnamefont
  {Evans}}, \bibinfo {author} {\bibfnamefont {D.}~\bibnamefont {Hinzke}},
  \bibinfo {author} {\bibfnamefont {U.}~\bibnamefont {Atxitia}}, \bibinfo
  {author} {\bibfnamefont {U.}~\bibnamefont {Nowak}}, \bibinfo {author}
  {\bibfnamefont {R.}~\bibnamefont {Chantrell}}, \ and\ \bibinfo {author}
  {\bibfnamefont {O.}~\bibnamefont {Chubykalo-Fesenko}},\ }\href {\doibase
  10.1103/PhysRevB.85.014433} {\bibfield  {journal} {\bibinfo  {journal} {Phys.
  Rev. B}\ }\textbf {\bibinfo {volume} {85}},\ \bibinfo {pages} {014433}
  (\bibinfo {year} {2012}{\natexlab{a}})}\BibitemShut {NoStop}%
\bibitem [{\citenamefont {Binder}(1997)}]{Binder1997}%
  \BibitemOpen
  \bibfield  {author} {\bibinfo {author} {\bibfnamefont {K.}~\bibnamefont
  {Binder}},\ }\href {http://iopscience.iop.org/0034-4885/60/5/001} {\bibfield
  {journal} {\bibinfo  {journal} {Reports Prog. Phys.}\ }\textbf {\bibinfo
  {volume} {60}},\ \bibinfo {pages} {487} (\bibinfo {year} {1997})}\BibitemShut
  {NoStop}%
\bibitem [{\citenamefont {Evans}\ \emph
  {et~al.}(2012{\natexlab{b}})\citenamefont {Evans}, \citenamefont {Chantrell},
  \citenamefont {Nowak}, \citenamefont {Lyberatos},\ and\ \citenamefont
  {Richter}}]{Evans2012a}%
  \BibitemOpen
  \bibfield  {author} {\bibinfo {author} {\bibfnamefont {R.~F.~L.}\
  \bibnamefont {Evans}}, \bibinfo {author} {\bibfnamefont {R.~W.}\ \bibnamefont
  {Chantrell}}, \bibinfo {author} {\bibfnamefont {U.}~\bibnamefont {Nowak}},
  \bibinfo {author} {\bibfnamefont {A.}~\bibnamefont {Lyberatos}}, \ and\
  \bibinfo {author} {\bibfnamefont {H.-J.}\ \bibnamefont {Richter}},\ }\href
  {\doibase 10.1063/1.3691196} {\bibfield  {journal} {\bibinfo  {journal}
  {Appl. Phys. Lett.}\ }\textbf {\bibinfo {volume} {100}},\ \bibinfo {pages}
  {102402} (\bibinfo {year} {2012}{\natexlab{b}})}\BibitemShut {NoStop}%
\bibitem [{\citenamefont {N\'{e}el}(1949)}]{Neel1949}%
  \BibitemOpen
  \bibfield  {author} {\bibinfo {author} {\bibfnamefont {L.}~\bibnamefont
  {N\'{e}el}},\ }\href@noop {} {\bibfield  {journal} {\bibinfo  {journal} {Ann.
  G\'{e}ophysique}\ }\textbf {\bibinfo {volume} {5}},\ \bibinfo {pages} {99}
  (\bibinfo {year} {1949})}\BibitemShut {NoStop}%
\end{thebibliography}%

\end{document}